# Scattering images from autocorrelation functions of P-wave seismic velocity images: the case of Tenerife Island (Canary Islands, Spain)

García-Yeguas, A. [1] [2] [7], Sánchez-Alzola, A. [3], De Siena, L. [4], Prudencio, J. [5] [2] [7], Díaz-Moreno, A. [8], Ibáñez, J. M. [6] [2]

(1) *Departamento de Física Aplicada, Universidad de Cádiz, Spain*
(2) *Instituto Andaluz de Geofísica. Universidad de Granada, Spain.*
(3) *Departamento de Estadística e Investigación Operativa, Universidad de Cádiz, Spain*
(4) *Departament of Geology and Petroleum Geology, University of Aberdeen, United Kingdom*
(5) *Earth and Planetary Science Department. University of California, Berkeley, USA*
(6) *Departamento de Física Teórica y del Cosmos. Universidad de Granada. Spain*
(7) *Instituto Volcanológico de Canarias (INVOLCAN). Tenerife, Spain*
(8) *Department of Earth, Ocean and Ecological Sciences. University of Liverpool. United Kingdom.*

\* Corresponding author. Tel.: +34956483320. e-mail address: araceli.garcia@uca.es

Abstract

We present a P-wave scattering image of the volcanic structures under Tenerife Island using the autocorrelation functions of *P*-wave vertical velocity fluctuations. We have applied a cluster analysis to total quality factor attenuation ($Q_t^{-1}$) and scattering quality factor attenuation ($Q_{PSc}^{-1}$) images to interpret the structures in terms of intrinsic and scattering attenuation variations on a 2D plane, corresponding to a depth of 2000 m, and check the robustness of the scattering imaging. The results show that scattering patterns are similar to total attenuation patterns in the South of the island. There are two main areas where patterns differ: at Cañadas-Teide-Pico Viejo Complex high total attenuation and average-to-low scattering values are observed. We interpret the difference as induced by intrinsic attenuation. In the Santiago Ridge Zone (SRZ) region, high scattering values correspond to average total attenuation. In our interpretation, the anomaly is induced by an extended scatterer, geometrically related to the surficial traces of Garachico and El Chinyero historical eruptions and the area of highest seismic activity during the 2004-2008 seismic crises.

1  Introduction

Since the second half of the Twentieth century different geophysical techniques have been developed to obtain direct images of the Earth's interior in a noninvasive way (e .g. seismic tomography (Aki et al. 1977); magnetotelluric (Cagniard 1953); gravimetry (Valliant 1991)). Among those employing travel times and amplitudes of seismic wave-packets, seismic tomography is the one that has evolved most over time, using different Earth properties as model parameters to represent the internal structure of the Earth and various types of observable signals as data (Aki et al. 1977; Nolet 2008).

The spatial scales resolved by using body-wave velocity tomography via measurements of P- and S-travel times can range from a few hundred meters in the crust (Patanè et al. 2017) to 500-1000 km in global tomography of the Earth mantle and core (Romanovicz, 2003; Nolet 2008). Seismic velocity tomography can be integrated by attenuation tomography, imaging the energy lost by P- and S-waves while traveling into the Earth (as quality factors Qp and Qs). With these techniques, researchers can better illuminate and interpret Earth features and characterize structural heterogeneities (Schurr et al. 2003; De Gori et al. 2005; Eberhart-Phillips et al. 2008). Attenuation is, in fact, more sensitive than velocity to strong spatial changes in the Earth composition, temperature, and pressure (Romanowicz, 2003) and an excellent marker of time-dependent changes in the physical and chemical properties of the medium (O'Connell and Budiansky, 1977; Fehler, 1982; Gusev and Lemzikov, 1985; Fehler et al., 1988).

At volcanic scale, integrated P- and S-wave velocity (Vp and Vs) and attenuation tomography models have therefore been increasingly successful at imaging volcanic structures (De Gori et al. 2005; De Siena et al. 2014; Prudencio et al. 2015a, 2015b). Still, we do not fully understand how seismic waves behave when interacting with extreme structural heterogeneity, thus greatly increasing the difficulty of modeling velocity and attenuation with deterministic physical theories and direct wave information (Sato et al.2012). Seismic coda waves (the late portion of an earthquake recording) have thus acquired increasing importance in active volcanoes to detect small-scale heterogeneity and magma plumbing systems (Sato et al. 2012; De Siena et al. 2014; De Siena et al. 2016). By using these secondary incoherent arrivals in the seismogram, total attenuation may be separated into intrinsic attenuation (energy dissipated by the seismic waves while travelling through the medium) and scattering attenuation (energy lost due to the interaction of the seismic waves with the medium heterogeneities, and effectively recorded later in the seismogram as coda). Several methods have been devised to separate and image these two contributions to total attenuation (Hoshiba et al., 2001; Akinci et al., 1995; Del Pezzo et al., 2001, 2006; Giampiccolo et al., 2006). In one of the first applications of scattering attenuation imaging, Nishigami (1997) found strong scattering bodies beneath volcanic areas, extending to 7 km (Mount Ontake volcano, Japan) and 20 km depths (Mount Nikko-Shirane volcano, Japan), respectively. The presence of a vertical region of high scattering under both volcanoes was interpreted as a magma conduit. Mikada et al (1997) found several patches of strong scatterers beneath Izu-Oshima volcano (Japan), which were interpreted as the primary magma reservoir. Revenaugh (1997) provides a review of the first scattering imaging studies applied to imaging geologic media. In the early 2000s, different techniques have been developed to obtain 2D and 3D intrinsic and scattering attenuation models for S-waves in volcanic areas (Frederiksen and Revenaugh 2004; Tramelli et al. 2006; Carcolé and Sato 2010; De Siena et al. 2014; De Siena et al. 2016). In particular, Prudencio et al. (2013) have used a diffusive approximation (Wegler and Lürh, 2001) to obtain intrinsic and scattering maps of Tenerife Island. To apply any of the above-mentioned methods it is necessary to use seismic waveforms.

When a P and S traveltime dataset is available but waveforms are too noisy to provide direct attenuation measurements, scattering attenuation measurements can be obtained using P- and S-wave velocity models. This method infers the scattering properties of P- and/or S-waves from three-dimensional velocity fluctuations, obtained from a velocity model (De Siena et al. 2011; Sato et al. 2012). The technique is based on the fact that, at scales smaller than a few hundred meters, velocity fluctuations tend to be random (Holliger et al. 1996). They can thus be obtained by using the spatial autocorrelation functions (ACF) of vertical velocity fluctuations calculated at different lateral points of a 3D velocity model (De Siena et al. 2011).

Among all possible statistical distributions, Von Karman and exponential distributions fit properly the velocity fluctuations. In fact, for the shape parameter λ=0.5 both distributions coincide. However, for exponential distribution the correlation length is directly related to the dominant characteristic scales lengths of the heterogeneities in the medium (Yoon, 2005). We need the correlation length (a) to make our interpretation and obtain scattering quality factor attenuation ($Q_{PSc}^{-1}$). It is today accepted the exponential distribution formodeling velocity fluctuations in either a stratified or volcanic medium, providing a statistical measure of the spatial scale and magnitude that characterizes heterogeneities in a volcano (Holliger et al. 1996; Shapiro and Hubral 1999, De Siena et al. 2013). In the Jemez volcanic field, spatial exponential ACFs of 3D velocity fluctuations are able to model strong stratification, image velocity changes when they are stronger in the vertical with respect to the horizontal direction, and identify a scattering anomaly beneath the volcanic cone (Sato et al. 2012). If used jointly with direct total attenuation measurements and interpreted via Cluster Analysis (CA), as at Campi Flegrei caldera (Italy) and Mount St. Helens volcano (USA), ACFs can be used to image and interpret scattering and intrinsic attenuation in 2D (De Siena et al. 2011, 2016).

Here, we have applied this technique to the island of Tenerife (Canary Islands, Spain), where models of velocity (García-Yeguas et al. 2012), total quality factor attenuation ($Q_t^{-1}$) (Prudencio et al. 2015a), and scattering and intrinsic attenuation of S-waves (Prudencio et al. 2013) are available. The aim of this work is to map $Q_{PSc}^{-1}$ at Tenerife Island in a 2D plane at ~2000m of depth and apply a CA to this and total attenuation measurements. We thus: (1) provide a new P-wave scattering image of the Island and (2) check the robustness of the method and interpret the results by comparing them with different seismic models. We have first calculated the ACFs of vertical velocity fluctuations measured from the P-wave seismic velocity model of Tenerife Island (Garcia-Yeguas 2010; Garcia-Yeguas et al. 2012) between depths of 100 m and 4500 m. The parameters controlling the ACFs are the correlation length (*a*) and the mean square fractional fluctuations ($\varepsilon^2$): these are used to calculate and map $Q_{PSc}^{-1}$ at a depth of ~2000 m assuming a travel time-corrected Born approximation (Aki and Richards 1980; Sato and Fehler 2012). Finally, we applied a CA to the images of $Q_{PSc}^{-1}$ (obtained in this work) and $Q_t^{-1}$ obtained by Prudencio et al. (2015a) at the corresponding depth, to interpret the scattering image and check the robustness of the results. The comparison of the results with the volcanic eruptive history, the known geological structures, and the recent seismic activity unravels important variations in the lateral structure and dynamics of the volcanic area.

## 2 Geological settings

Tenerife Island is a volcanic island that belongs to the Canary Islands archipelago (Spain) (Figure 1). It is the major island of the Canary Islands with an area of 2,040 km$^2$. Its geomorphology is very heterogeneous, with monogenetic cones, basaltic and felsic lava flows, stratovolcanoes, domes, and pyroclastic deposits; evidence of diverse volcanic processes (Ablay et al. 1995; Dóniz et al. 2008; Dóniz-Páez 2015 95 Romero 1991, 1992).

The most significant volcanic complex is located in the center of the island and comprises Las Cañadas caldera, an elliptical 16 × 9 km caldera lying at 2000 m above sea level (LCC in figure 1), within which lies a stratovolcanic complex identified as Teide-Pico Viejo (Teide-PV in figure 1). Together these structures form the Cañadas-Teide-Pico Viejo Complex (CTPVC in figure 1).

Different periods of activity have been identified in the evolution of CTPVC evolution, separated by longer intervals of quiescence (Araña et al. 1994). CTPVC has two rift zones: the Santiago Rift Zone (SRZ in Figure 1) to the NW and the Dorsal Rift Zone (DRZ in Figure 1) to the NE. The last historical eruptions took place, with the exception of the Chahorra eruption, that occurred in 1798 within the CTPVC complex, all historical eruptions have taken place along these two rift zones (Romero 1991, 1992). Sietefuentes (1704), and Fasnia and Arafo (1705), and the SRZ eruptions have been Garachico volcano (1706) and El Chinyero (1909).

Several studies have imaged the inner structure of Tenerife Island using different physical properties: resistivity measurements (Pous et al. 2002; Coppo et al. 2008; Piña-Varas et al. 2014; Piña-Varas et al. 2015); aeromagnetic surveying (Blanco-Montenegro et al. 2011), gravity (Araña et al. 2000; Gottsmann et al. 2008) and seismological studies (Canales et al. 2000; García-Yeguas 2010; García-Yeguas et al. 2012; Prudencio et al. 2013, 2015a).

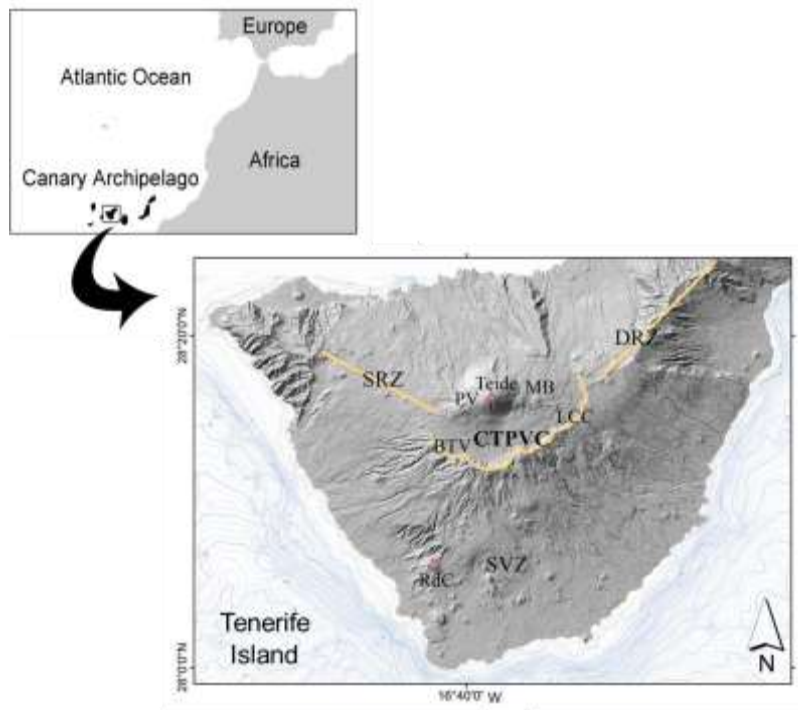

Fig. 1: Regional setting and location of Tenerife Island. Rift position and Las Cañadas wall are marked with yellow lines. CTPVC: Cañadas-Teide-Pico Viejo Complex, SVZ: Southern Volcanic Zone, LCC: Las Cañadas Complex, PV: Pico Viejo, MB: Montaña Blanca, BTV: Boca Tauce Volcano, SRZ: Santiago Rift Zone, RdC: Roque del Conde, DRZ: Dorsal Rift Zone and FASF: Fasnia-Arafo-Siete Fuentes eruption.

3 Data and Methodology

To perform this study we have used the high-resolution 3D P-wave seismic velocity and attenuation models of Garcia-Yeguas et al. (2012) and Prudencio et al. (2015a). Both models were obtained using the seismic recordings of an active seismic experiment (TOM-TEIDEVS - Ibáñez et al. 2008). More than 6300 shots were fired by air guns connected to a BIO Hespérides research vessel and recorded at 125 seismometers deployed onland. A subset of 103,750 high-quality travel times and waveforms were chosen to perform the tomographic inversions. The velocity model has lateral extension of $40 \times 40$ km$^2$ and depth ranging from the top of Teide volcano (over 3000 m above sea level (a.s.l.)) to 8000 m below sea level (b.s.l.) at a node spacing of 0.7 km in all directions.

To obtain the ACF of vertical velocity fluctuations and derive a $Q_{PSc}^{-1}$ image at a depth of ~2000 m we have selected the regions of maximum resolution and reliability between depths of 0.1 km and 4.5 km, following the resolution and reliability tests of the two tomographic studies (see Garcia-Yeguas et al. (2012) and Prudencio et al. (2015a) for details). We have applied a CA using as input parameters the values of $Q_{PSc}^{-1}$ and the values of $Q_t^{-1}$ obtained by Prudencio et al. (2015a) at a depth of 2000 m.

The magnitude and scale of random heterogeneities in a volcanic medium can be quantified by the ACF of velocity fluctuation distributions (Aki and Chouet 1975). The $a$ and the $\varepsilon^2$ control the shape and amplitude of the ACF and are sensitive to the physical state of the upper crust (Sato and Fehler 2012). Here, we have followed the methodology described by De Siena et al. (2011), using as data the 3D velocity measurements obtained by García-Yeguas et al. (2012). These data are distributed on a regular grid of 0.7 × 0.7 × 0.7 km step, providing seven vertical high-resolution velocity measurements at each point on the 2D map. We fit a $N^{th}$-order average polynomial over the seven vertical measurements at each point on the map to remove the vertical trend and randomize the velocity values. We determine the polynomial degree of the generic velocity function as:

$$V(z) = a_0 + a_1 z + a_2 z^2 + \ldots + a_N z^N \quad (1)$$

This is obtained using the Bayesian (or Schwartz, 1978) Information Criterion (BIC) and considering equal variance values $\hat{\sigma}_e^2$ at all nodes:

$$\text{BIC} = n\ln(\hat{\sigma}_e^2) + N\ln(n) \quad (2)$$

where $\hat{\sigma}_e^2$ is the error variance for normally distributed errors, N is the number of model parameters to be obtained (polynomial degree) and $n$ the number of observations. A higher degree of the polynomial is better to fit the trend, but the BIC criterion avoids over-fitting introducing a penalty term against the increasing number of parameters. After averaging the BIC results over the entire resolved map, and using the elbow method (Hartigan 1975), we observe that a first order polynomial (or a simple regression) with two parameters:

$$V(z) = a_0 + a_1 z \quad (3)$$

This provies a best fit the vertical trend at best. Figure 2a shows the BIC as a function of the polynomial degree. We observe that polynomial degrees greater than one do not improve significantly the BIC values.

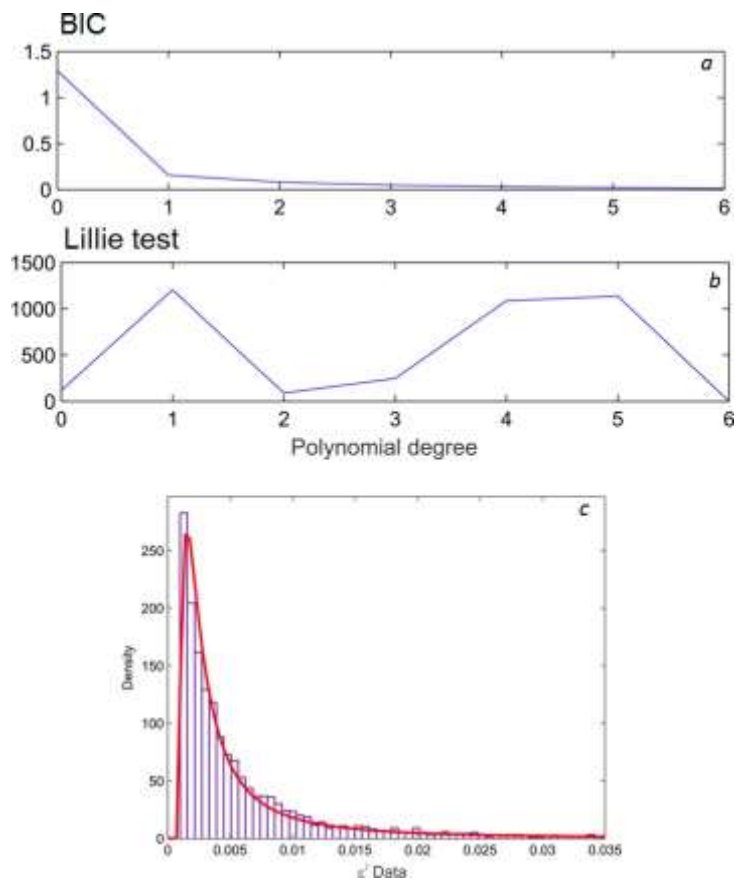

Fig. 2: a. The BIC criterion as a function of polynomial order. b. Number of de-trended random Gaussian measurements on the map showing random fluctuations after de-trending and a Lilliefors test (Lilliefors, 1967) c. Histogram of the 2D $\varepsilon^2$ spatial variations fitted with a Generalized Extreme Value distribution (GEV - red line).

In a perfectly random medium the velocity fluctuations should be random after de-trending. We have applied the Lillie test (Lilliefors, 1967) at 5% confidence to the velocity fluctuations at all points on the map after de-trending. The null hypothesis is that the data distribution is normal. For different polynomial degrees, we store the points having a p-value (probabilities that support the null hypothesis) greater than the significance level of 0.05, where the null hypothesis is not rejected. In figure 2b we plot the number of points that passed the test against the order of the polynomial used for de-trending. The highest number of points presenting random Gaussian fluctuations corresponds to the lowest degree in the polynomial (first order polynomial). Our aim is to obtain the maximum number of random points using the polynomial with the lowest polynomial degree N. The values of Lillie test for N=1 and N=4-5 shown in Fig 2 are fairly similar, with maximum number of blocks around 1200-1300; however, only polynomials of N = 1 and N = 2 provide comparable values for the BIC. If we had obtained a similar amount of random blocks with N=1 and N =2 with the Lillie test, our choice of a polynomial of order 1 would have been unjustified. As there is an increase in number of random blocks only for order N=4, the best choice is N=1. The de-trend of the velocity model in the area is thus done using best-fit coefficients for the polynomial:

$$\tilde{v}(z) = V^*(z) - a^*_0 - a^*_1(z) \qquad (4)$$

where $\tilde{v}(z)$ are the velocity fluctuation distribution with depth for any surface grid point ($x*$, $y*$). The normalized ACF, $R(\Delta z)$, is calculated numerically as:

$$R(z) = \langle \tilde{v}(z)\tilde{v}(z + \Delta z) \rangle \qquad (5)$$

where the brackets indicate spatial average and $\Delta z$ is the correlation space-lag positive for the increasing depth. Additionally, we have used an exponential ACF model to adjust the random velocity fluctuation (Shapiro and Hubral, 1999). We show some examples of ACF values and its fit in the Supplementary Material (figure S1). This model depends on $\varepsilon^2$ and $a$ as:

$$R(z) = \varepsilon^2 exp(-z/a) \qquad (6)$$

where the maximum of $R(z)$ coincides with $\varepsilon^2$ at $z = 0$. We compute the values of space averaged $a$ as being 1.968 ± 0.013 km and the space-averaged value $\varepsilon^2$ is 0.00365 ± 0.00015 (these values have been computed with a confidence of 95%). We also fit a Generalized Extreme Value distribution (Fréchét distribution) to the $\varepsilon^2$ values obtained at all grid points on the map. This distribution has been shown to fit the velocity fluctuations as a sequence of non-gaussian independent and identically distributed random variables and in the presence of strong heterogeneity (De Siena et al. 2011; De Siena et al. 2016). Figure 2c shows the fit of the 2D $\varepsilon^2$ with a generalized extreme value distribution with a shape parameter $\lambda$ of 0.000815. As De Siena et al. (2011), we remark that the $a$ and the $\varepsilon^2$ are biased. We also note that the $a$ measurements are affected by larger uncertainties.

Both $\varepsilon^2$ and $a$ parameters are used to model $Q^{-1}$ scattering for P-waves ($Q_{PSc}^{-1}$). We have followed the theory of elastic random fluctuations in the Born approximation assuming an exponential ACF. This approximation considers a cut-off wavenumber value ($v_c$) $v_c^2 = ¼$ to correct high-frequency dependencies at a determined wave number ($\kappa$) (De Siena et al., 2011; Sato and Fehler, 2012).

$$Q_{PSc}^{-1} = \frac{15\varepsilon^2 a^3 \kappa^3}{2(1+\frac{a^2\kappa^2}{4})(1+4a^2\kappa^2)} \qquad (7)$$

where $\kappa = 2\pi f/v$, f is frequency in Hz and v is the P-wave seismic velocity in km/s. In this study we have chosen f = 6 Hz as this is the central frequency used to filter the seismic signals to perform the velocity tomography (García-Yeguas et al. 2012), and we have assumed v = 6 km/s as the average P-wave seismic velocity of the model (García-Yeguas et al. 2012). Finally, we have applied a K-means cluster analysis method to relate in space the variations of the $Q_{PSc}^{-1}$ and with those of the total quality factors ($Q_{Pt}^{-1}$) (Hartigan 1975; De Siena et al., 2011, 2016). On the other hand, as we are only interested in areas of high and low $Q_{PSc}^{-1}$ and $Q_{Pt}^{-1}$ (two options for each variable), picking four clusters makes interpretation of the anomalies easier. This is only true if the difference in reduction in point-to-centroid distance of four clusters with respect to three or five is small, as in our case. Previously we have made a statistical analysis comparing both variables to check unreliable spatial measurements of $Q_{PSc}^{-1}$ and $Q_{Pt}^{-1}$ dispersion. This study is included in the Supplementary Material (figure S2). We have used the Euclidean distance to group the two different variables in their parameter space: $Q_{PSc}^{-1}$ obtained in this study, and $Q_{Pt}^{-1}$ measured by Prudencio et al. (2015a) at 2000 m b.s.l. To choose the number of clusters, we plot the percentage reduction in point-to-centroid distance against the number of clusters (Figure 3a). The ideal number of clusters is that after which the distance is not consistently improved increasing number of clusters i. e. the point after which the curve starts a plateau and/or goes over a set threshold (80%). The elbow method suggests the need of K = 4 clusters, at the start of the plateau of percentage reduction and where we achieve almost 80% reduction of the cluster-to-observation distance (De Siena et al. 2011; De Siena et al. 2016). Figure 3b shows the variables (dots) in their parameter space, labeled with a color corresponding to each cluster. A crossed circle marks the cluster centroids. We observe that red plots are in regions where scattering attenuation is larger than total attenuation. This is due to the use of two different methodologies to obtain total attenuation — Prudencio et al.(2015) and scattering attenuation (De Siena et al., 2011). The inverse total quality factor cannot be retrieved by the sum of the scattering and intrinsic quality factor exactly with this analysis, as stated in De Siena et al. (2011). The results can only be interpreted qualitatively using clusters analysis. Furthermore, the K-means algorithm also classifies the data by taking into account the point-to-centroid distance only, without taking into account other aspects (such as geological formations and the inner structure of the volcano) which could better constrain our analysis.

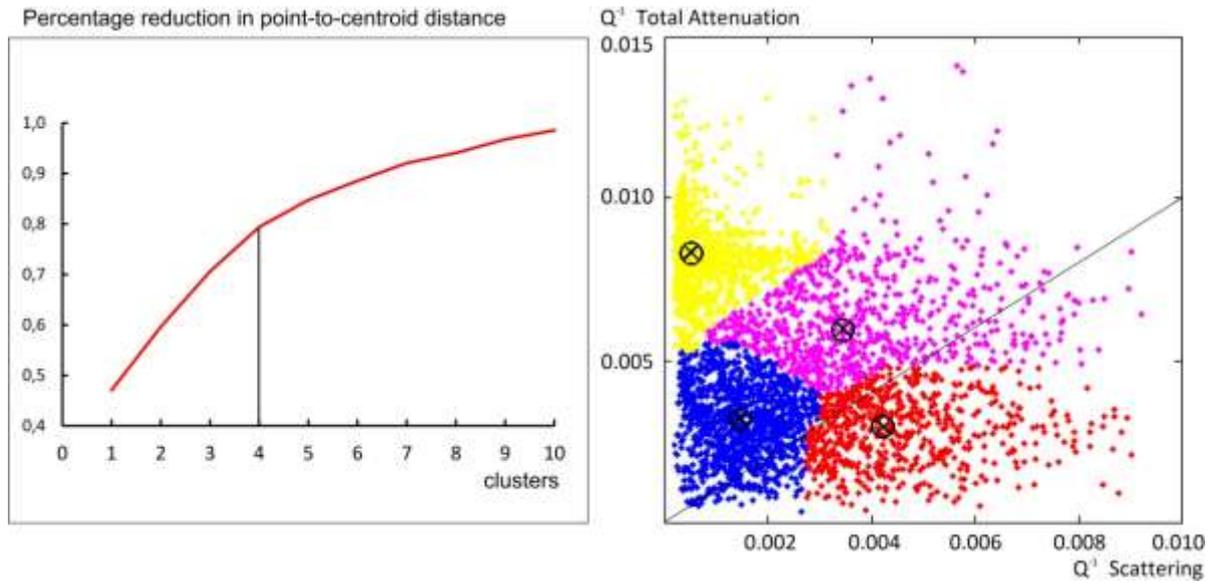

Fig. 3: a. Percent reduction in point-to-centroid distance with respect to the number of clusters. The black line shows the 80% reduction obtained with 4 clusters. b. Results of the 4-means CA . A black-crossed circumference marks the centroids. The black line divides the image in two sections. The points located in the lower section are considered erroneous because $Q_{PSc}^{-1}$ is greater than $Q_{Pt}^{-1}$. The percentage of erroneous points is 10.94 % in the studied area (from the original grid 40 km × 40 km). The most affected cluster is the red one. We assume these regions are affected by anomalously high scattering.

4  Results and discussion

Based on the checkerboard tests carried out by García-Yeguas et al. (2012), we have chosen only the region of the island above sea level, where the resolution is higher in this area. To provide a first qualitative interpretation of the scattering image, and check the robustness of the method, we compare the $\varepsilon^2$ and $Q_{\bar{P}Sc}^{-1}$ images with the P-wave total attenuation model (Prudencio et al. 2015a). The *a* parameter values vary from ~0.37 km to a maximum of 2.91 km. However, De Siena et al. (2016) pointed out that this variable is influenced by huge uncertainties; therefore, we consider $\varepsilon^2$ as the best parameter to model P-waves heterogeneities at Tenerife Island.

4.1 Spatial patterns of $\varepsilon^2$ and $Q_{PSc}^{-1}$

In figure 4 we show the 2D measurements of $\varepsilon^2$ and derived $Q_{PSc}^{-1}$ super-imposed on a contour-map of Tenerife Island. The spatial patterns of $\varepsilon^2$ identify areas of high scattering heterogeneity and control seismo-volcanic coda envelope shapes at intermediate and late lapse times (De Siena et al. 2013). Figure 4a shows high $\varepsilon^2$ values in the Santiago Rift Zone (SRZ) and medium and low values across the rest of the island.

The highly-heterogeneous $\varepsilon^2$ patterns in the SRZ are spatially correlated to historical eruptions in this area (Romero 1991, 1992) including Garachico (occurred in 1706) and El Chinyero (in 1909, the last eruption occurred in Tenerife). The $\varepsilon^2$ values, related to the mean squared fractional fluctuation, are estimated in this study as varying from 0.00 to 0.12. These values are also consistent with other volcanic studies such as De Siena et al. (2011). We consider that high values of $\varepsilon^2$ could be related to either the presence of non-consolidated and fragmented materials or residual hot magmatic intrusions, as they generally correspond to low P-wave velocities (García-Yeguas et al. 2012) and density values (Gottsmann et al. 2008). The seismicity recorded between 2004 and 2008 is the expression of a general volcanic reactivation and mainly affects NW sector of the island (Almendros et al. 2007; Cerdeña et al. 2011). While high $\varepsilon^2$ in this region corresponds to the presence of cinder cones, historical eruptions, and recent seismicity (figure 4a), the relationship of high $\varepsilon^2$ to geological heterogeneities and seismic activity is unclear in the rest of the island.

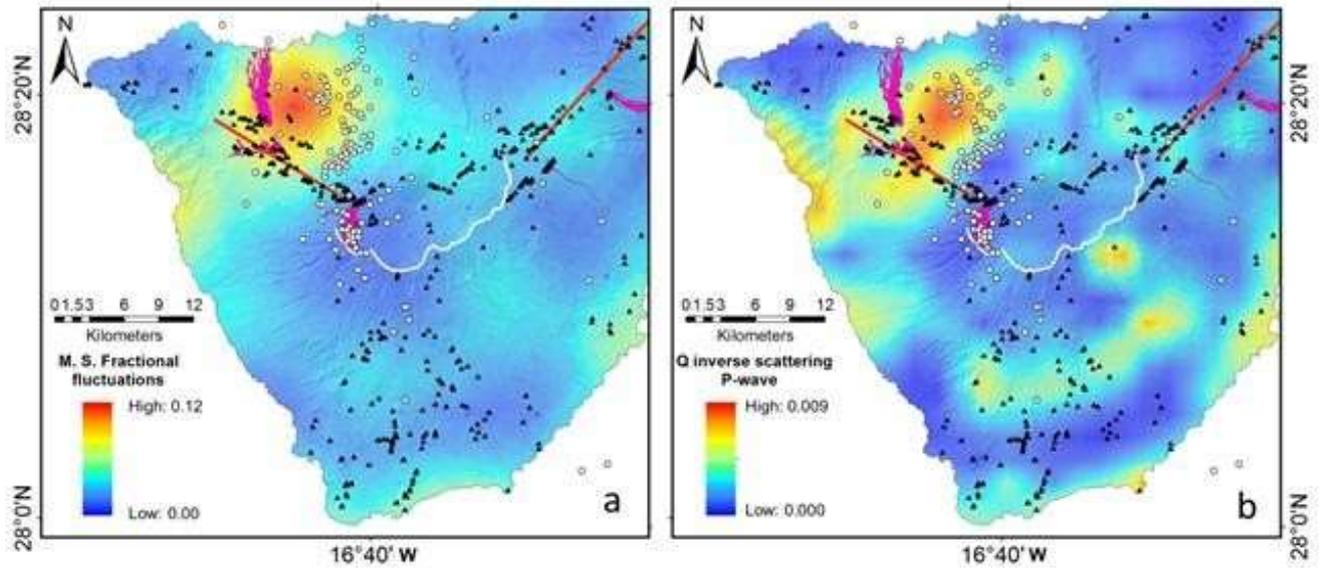

Fig. 4: a. Spatial patterns of $\varepsilon^2$ measured from the ACF distribution. b. P-wave $Q^{-1}$ scattering. White circles indicate the epicenters of the earthquakes between 2004 and 2008 (from Cerdeña et al. 2011). Black triangles and purple areas mark cinder cones and historical eruptions, respectively. Red lines indicate SRZ (left) and DRZ (right). The white line outlines the rim of Las Cañadas Caldera.

The map of inverse quality factor for P-waves ($Q^{-1}_{PSc}$, figure 4b) highlights regions showing different scattering characteristics: i) the SRZ, southeastern and southern edges of the island are characterized by the highest values of $Q^{-1}_{PSc}$, ii) DRZ and southeastern parts of the island display average $Q^{-1}_{PSc}$ values, and iii) the central, south-central, and north-easternmost parts of the island show low $Q^{-1}_{PSc}$ values. High values of $Q^{-1}_{PSc}$ at the SRZ are correlated with low P-wave velocity and density values at depth (Gottsmann et al. 2008; García-Yeguas et al. 2012). While these results hint atfluid and melt accumulation at depth, the region is also characterized at the surface by unconsolidated products of the last historical eruptions in

Tenerife (Garachico in 1706 and El Chinyero in 1909), which may affect coda recordings at frequencies characteristic of local earthquake tomography (De Siena et al. 2016). Prudencio et al. (2013) obtained $Q^{-1}$ S-wave scattering attenuation maps from the inversion of the energy envelopes recorded at different source–receiver pairs, showing a similar--although damped--high-scattering anomaly in the area. East of the anomaly, the seismicity that occurred between 2004 and 2008 matches with high $Q_{PSc}^{-1}$ values in this region. The hypocenters of the earthquakes in Northwest Tenerife are mainly located between depths of 0 and 6 km (Cerdeña et al. 2011) and our map is produced for a depth of ~2000 m using data from 0.1 km to 4.5 km depths. We relate the seismicity with an unconsolidated area in the North of the island, and high values at the southeastern and southern edges of the island could be related to fragmented or/and unconsolidated material at these locations.

The DRZ and Southeast of Tenerife display medium-high values of $Q_{PSc}^{-1}$. We associate the anomalies in the DRZ with unconsolidated material of the last eruptions in 1704 (Siete Fuentes) and 1705 (Fasnia). In the southeastern part of the island there has been no recent volcanic activity, thus we associate this $Q_{PSc}^{-1}$ values with volcanoclastic deposits and fragmented material. On the other hand, the central, south-central, and north-easternmost parts of the island have low $Q_{PSc}^{-1}$ values. We relate these areas to consolidated cool non-fragmented bodies. This interpretation is in agreement with the presence of a high-velocity (García-Yeguas et al. 2012) and high-density (Gottsman et al. 2008) body in the center of Tenerife.

4.2. Cluster analysis (CA) results

Figure 5c shows the results obtained from the CA using the values of $Q_{SC}^{-1}$ obtained in this study (figure 5a) and the total attenuation tomography model provided by Prudencio et al.

(2015a) (figure 5b). The result provides a quantitative way to interpret attenuation in terms of intrinsic and scattering mechanisms on a 2D plane corresponding to a depth of ~2000 m. The CA patterns show (figure 5c):

- Regions in pink (high $Q^{-1}_{PSc}$ and high $Q^{-1}_t$): these regions are characterized by high values of both $Q^{-1}_{PSc}$ and $Q^{-1}_t$. In Figure 5c, the pink areas located outside of Las Cañadas were identified as high intrinsic attenuation regions by Prudencio et al. (2013). Following Prudencio et al. (2013), we interpret these anomalies has being induced by fragmented materials and/or volcanoclastic deposits.

- Regions in blue (low $Q^{-1}_{PSc}$ and low $Q^{-1}_t$): Neither scattering nor intrinsic attenuation dominate these areas. We infer these regions are mainly comprised of well consolidated materials (Del Pezzo, 2008; Prudencio et al. 2017).

- Regions in yellow (low $Q^{-1}_{PSc}$ and high $Q^{-1}_t$): In these regions, P-wave attenuation is mainly due to intrinsic mechanisms. Some authors (Mayeda et al. 2008; Prudencio et al. 2013, for example) highlight that attenuation in volcanic regions is generally produced by scattering processes. The intrinsic absorption dominates over the scattering attenuation for regions where low seismic activity occurs. We observe that these regions are located in the outer part of the island and there is not seismic activity in them (Canas et al. 1998).

- Regions in red (high $Q^{-1}_{PSc}$ and low $Q^{-1}_t$): Northeast of CTPVC, the SRZ and DRZ are all characterized by anomalously high scattering attenuation. In figure 3b we can observe that most of the attenuation values in this cluster are mainly due to scattering, in comparison with the pink cluster. De Barros et al. (2012) identified these regions as being associated with ancient magmatic bodies which fed eruptions of the last 2000 years in CTPVC. In addition, high values of scattering and measurable E-W compression (Sánchez-Alzola et al. 2016) can be related to the presence of fragmented materials mixed with consolidated bodies. They maybe related to dikes and ancient conduits of the last historic volcanic eruptions (Boca Cangrejo, 1492; Garachico, 1706; Chinyero, 1909). In the South of Tenerife Island there are also small patterns of high $Q^{-1}_{PSc}$ and low $Q^{-1}_t$. As there is no evidence of recent volcanic activity, these anomalies are more feasibly related to old volcanoclastic deposits combined with ancient consolidated constructs.

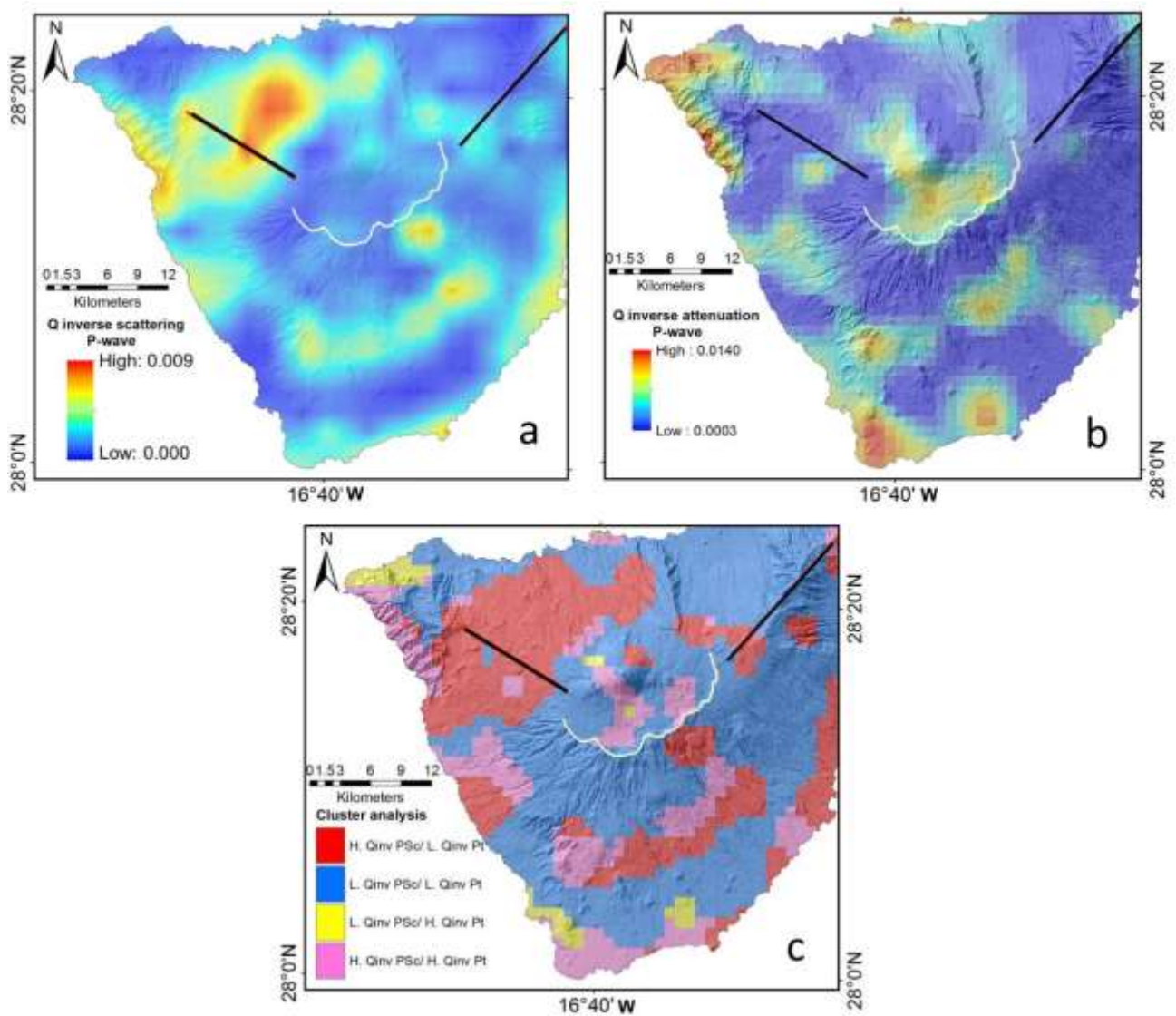

Fig. 5: a. $Q^{-1}_{PSc}$ map, b. $Q^{-1}_{t}$ map, and c. CA distribution. Red: High $Q^{-1}_{PSc}$ and low $Q^{-1}_{t}$. Blue: Low $Q^{-1}_{PSc}$ and low $Q^{-1}_{t}$. Yellow: Low $Q^{-1}_{PSc}$ and high $Q^{-1}_{t}$. Pink: High $Q^{-1}_{PSc}$ and high $Q^{-1}_{t}$. Black lines indicate SRZ (left) and DRZ (right). White line designates Las Cañadas Caldera wall.

5 Conclusions

In this work, a P-wave scattering ($Q^{-1}_{PSc}$) map has been obtained from the vertical autocorrelation functions (ACF) of measured spatial velocity variations. The $Q^{-1}_{PSc}$ distribution is obtained by using the measured $a$ and $\varepsilon^2$ in a single scattering approximation. The method is particularly valid when seismograms are either unavailable or difficult to read. The results confirm that the acquisition of a scattering image from a tomographically defined velocity model increases our ability to interpret active volcanic structures. The CA technique is applied to scattering and total attenuation anomalies to separate scattering and intrinsic attenuation on a 2D map at a depth of ~2000m quantitatively; the comparison with literature and volcanological information shows the robustness of the technique, particularly in retrieving high-scattering, magma-related anomalies.

Focusing on $\varepsilon^2$ and a, we note that high $\varepsilon^2$ values can be associated with high correlation length values (De Siena et al. 2011). The scattering distribution is similar to the total attenuation distribution in the South of the island. In other parts of Tenerife, we observe two alternative settings. First at CTPVC low/average scattering attenuation corresponds to high total attenuation. We infer from these values that the caldera is filled by unconsolidated rocks and fragmented and fractured material from Teide–Pico Viejo stratovolcanoes (Coppo et al. 2008; Coppo et al. 2010; Piña-Varas et al. 2014; Prudencio et al. (2015a); and Villasante-Marcos et al. 2014). Second at SRZ the high scattering attenuation is paired with low/average total attenuation. We relate these characteristics to the existence of a strong extended scatter, linked with past volcanic (historical eruptions: Garachico and El Chinyero) and seismic activity (seismic crisis of 2004-2008).

Acknowledgements

We thank Edoardo Del Pezzo for the valuable idea of this paper and suggestions regarding the methodology. J. Prudencio is partially supported by NSF1521855 Hazard SEES project. This paper has been partially supported by the Spanish project KNOWAVES (TEC2015-68752-R (MINECO/FEDER)), the European project MED-SUV funded by the European Union's Seventh Framework Program for research, technological development and demonstration under Grant Agreement No 308665, and by the Regional project 'Grupo de Investigación en Geofísica y Sismología de la Junta de Andalucía, RNM104'.